\documentclass[iop,apj]{emulateapj}

\usepackage{graphicx}	
\usepackage{amsmath}	
\usepackage{amssymb}	
\usepackage{units}
\usepackage{enumitem}
\usepackage{bm}
\usepackage{color}
\usepackage[breaklinks,colorlinks,citecolor=blue,linkcolor=red]{hyperref} 

\def\msun{{\rm\,M_\odot}}

\def\msun{{\rm\,M_\odot}}

\newcommand{\kms}{\, {\rm km\, s}^{-1}}

\newcommand{\be}{\begin{equation}}
\newcommand{\ee}{\end{equation}}

\def\h2{${\rm\,H_2}$}

\newcommand{\beq}{\begin{equation}}
\newcommand{\beqa}{\begin{eqnarray}}
		 \newcommand{\eeq}{\end{equation}}
\newcommand{\eeqa}{\end{eqnarray}}

\begin{document}
\title{Explaining the LIGO black hole mass function with field binaries: Revisiting Stellar Evolution at low Metallicity or Invoking Growth via gas accretion? }

\author{Mohammadtaher Safarzadeh\altaffilmark{1,2} \& Enrico Ramirez-Ruiz\altaffilmark{2}}
\affil{$^1$Center for Astrophysics | Harvard \& Smithsonian, 60 Garden Street, Cambridge, MA; \href{mailto:msafarzadeh@cfa.harvard.edu}{msafarzadeh@cfa.harvard.edu}}
\affil{$^2$Department of Astronomy \& Astrophysics, University of California, Santa Cruz, CA 95064, USA}

\begin{abstract}
Our understanding of the formation and evolution of binary black holes (BBHs) is significantly impacted by the recent discoveries made by the LIGO/Virgo collaboration. Of utmost importance is the detection of the most massive BBH system, GW190521.
Here we investigate what it takes for field massive stellar binaries to account for the formation of such massive BBHs. Whether the high mass end of the BH mass function is populated by remnants  of massive stars that either formed at extremely low metallicities and avoid the pair-instability mass gap or increase their birth mass beyond the pair-instability mass gap through the accretion of gas from the surrounding medium. We show that assuming that massive stars at very low metallicities can form massive BHs by avoiding pair-instability supernova, coupled with a correspondingly high formation efficiency for  BBHs, can explain the observed BH mass function.
To this end, one  requires a relation between the initial and final mass of the progenitor stars at low metallicities that is shallower than what is expected from wind mass loss alone. On the other hand, assuming pair-instability operates at all metallicities, one can account for the observed BH mass function if at least about 10\% of the BHs born at very low metallicities double their mass before they merge because of accretion of ambient gas. Such BBHs will have to spend about a Gyr within a parsec length-scale of their parent atomic cooling halos or a shorter timescale if they reside in the inner sub-parsecs of their host dark matter halos. Future stellar evolution calculations of massive stars at very low metallicity  and hydrodynamical simulations of gas accretion onto BBHs born in atomic cooling halos can shed light on this debate. 
\end{abstract}

\section{Introduction}
During the third observing run of LIGO/Virgo collaboration (LVC), numerous unexpected binary black hole (BBH) merger events were detected that shook the scientific community. 
A particularly notable example is the discovery of the most massive BBH merger, GW190521, with a total mass of 150 $\msun$, which has challenged  our understanding of the BH mass function \citep{Abbott:2020dz,Abbott:2020gz}. 
Before this detection, the maximum mass was estimated to be around 45 $M_{\odot}$ \citep{Roulet2019,Perna2019}, consistent with the predictions of the pair-instability supernova \citep{Woosley:2017dj,Spera2017,Marchant2019,Stevenson2019}. With the detection of BHs in the pair-instability mass gap in the third observing run of the LVC \citep{O3a}, the BH mass function is now estimated to have a maximum mass of about 90 $M_{\odot}$.  

The many astrophysical theories put forward  to explain the origin of this even can be categorized  into three broad groups\footnote{Non-astrophysical model proposals include, for example, changes in nuclear reaction rates to physics beyond the Standard Model \citep{Sakstein2020}.}: massive BBHs can form from progenitor stars at extremely low metallicities that are speculated to avoid the pair-instability supernova mass limit \citep{Farrell+2020, Kinugawa+2020};  massive BBHs growing via  accretion of surrounding gas  \citep{McKernan+2012,Tagawa2016,Tagawa:2019tu,Yang+2019,Roupas2019,Luca2020,SH2020,Liu2020ApJ};
 and massive BBHs formed as a product of hierarchical mergers of smaller mass BHs ~\citep{Perna2019,Fragione:2020uh,Rodriguez2020,Rizzutto+2020}.

In this paper, rather than focusing on solely explaining the nature of GW190521, we turn our attention to the origin of the merging BBH mass function deduced in the first half of the third observing run of the LVC. 
The massive end of the newly derived mass function does not fall steeply after about 40 $\msun$ but now extends to about 100 $\msun$. In particular we investigate whether relaxing the pair-instability limit at low metallicities or inducing ambient gas accretion can help explain the observed merging BBH mass function. We do not consider the possibility of other BBH formation channels  such as hierarchical mergers as we center our attention to the constraints that can be derived  for field massive stellar binaries. 
We show that both scenarios are capable of explaining the observed mass function with caveat for each scenario that is thoroughly discussed in this work. 

The structure of this work is as follows: in \S2 we describe or analytic method to estimate the BBH formation budget and merging rate as a function of redshift for different metallicity bins. In \S3 we compare the results of the predicted mass function in the scenario in which the pair-instability is relaxed at low metallicities and compare our results with those obtained from population synthesis studies. In \S4 the accretion channel results are discussed and compared to the other approach. Finally, in \S5, we present our conclusions. 

\section{Method}
\subsection{Star formation history, metallicity evolution and their parametrization} 
The star formation rate sets the global budget for the formation of compact binary objects. It has been determined through UltraViolet (UV)-optical, Near Infrared (NIR), far-infrared (FIR), and sub-millimeter observations of the galaxies across the cosmic time. 
It is parametrized as \citep{Madau:2014gtb}:
\be
\psi(z)=0.015 \frac{(1+z)^{2.7}}{1+[(1+z)/2.9]^{5.6}}\,\, \msun\, {\rm yr^{-1}\, Mpc^{-3}}.
\ee 

\begin{figure}
\hspace{-0.2in}
\centering
\includegraphics[width=\columnwidth]{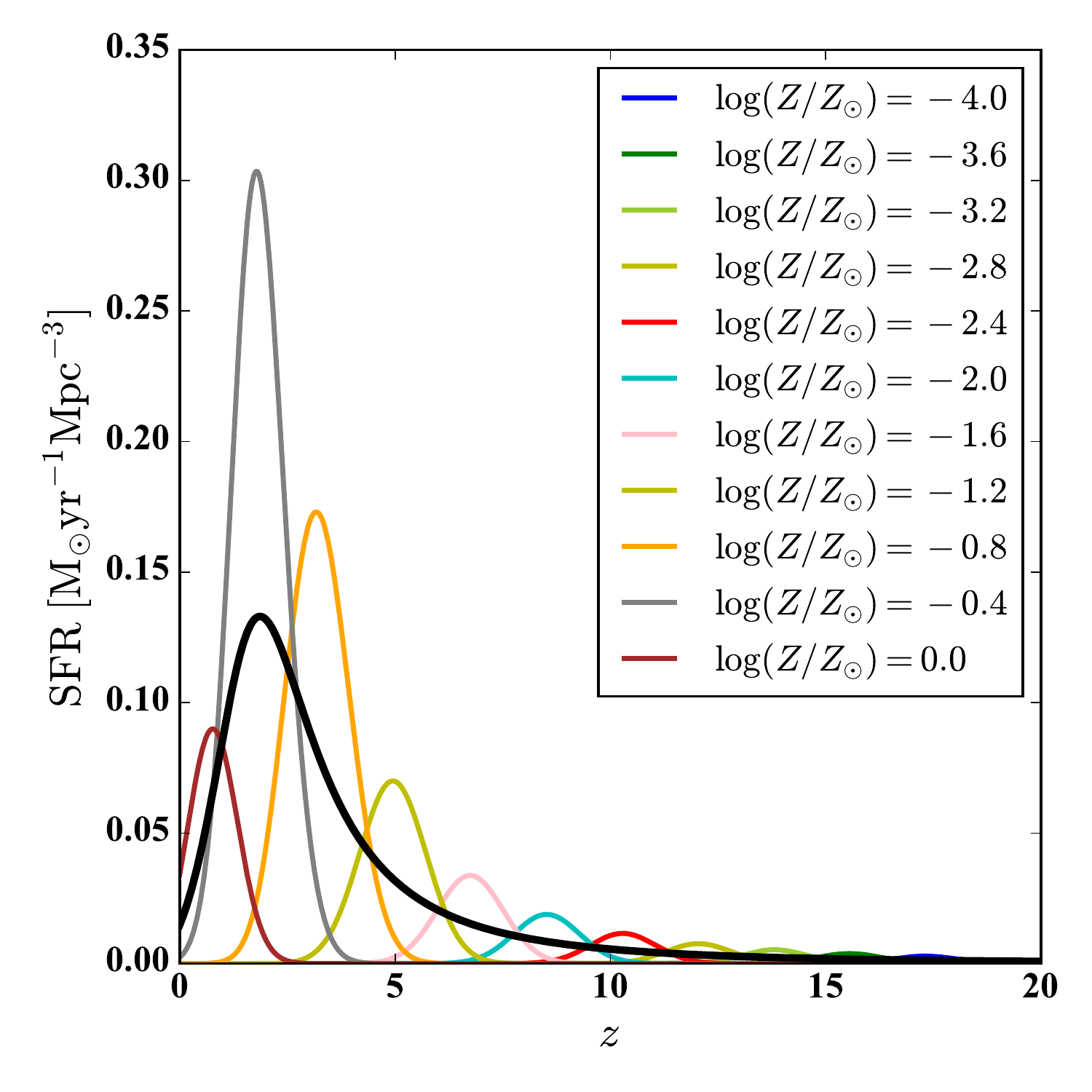}
\caption{The overall SFR history of the universe shown in black solid line. The colored lines show the SFR history at a given metallicity bin.}
\label{fig:SFR_Z}
\end{figure}

One crucial piece of information is the fraction of the star formation that is taking place at a specific metallicity range. This, as we will see shortly, affects the mass of the BBHs. 
Since our knowledge of the metallicity evolution of the universe is rather incomplete, there is not a well-defined way to parametrize it. 
Here we adopt a commonly used  heuristic approach so that  
\be
\frac{d\psi(z)}{d\log Z}=\psi(z) \frac{1}{\sigma\sqrt{2\pi}} e^{-\frac{\ln(Z)-\mu(z)}{2\sigma^2}},
\ee
where $\mu(z)$ defines the mean metallicity at a given redshift parametrized as:
\be
\mu(z)=Z_0\times10^{\alpha z},
\ee
with $\alpha=-0.23$, $Z_0=1.0$, and $\sigma=0.4$ \citep{Neijssel2019}. Figure \ref{fig:SFR_Z} shows the overall SFR history in black and in color shows the SFR history for a given metallicity bin.

Two compact objects in a binary with orbital semi-major axis $a$ merge on a timescale \citep{Peters1964},
\be
\tau=\frac{5}{256} \frac{c^5}{G^3}\frac{a^4}{(m_1 m_2)(m_1+m_2)},
\ee
where $m_1$ and $m_2$ are the mass of the primary and secondary compact objects. 
The binary's orbit shrink due to the emission of GWs, which carry away both energy and angular momentum from the binary. The distribution of compact object binaries' semi-major axis at birth sets the timescale over which they merge. 
Assuming the distribution of the binaries in semi-major axis ($a$) follows 
\be
\frac{dN}{d a}\propto a^{-\nu},
\ee
and given that the merging timescale due to emission of gravitational waves scales as $\tau\propto a^{4}$, the distribution of merging timescales would be given by:
\be
P(\tau)\equiv\frac{dN}{d\tau}=\frac{dN}{da}\frac{da}{d\tau}\propto a^{-\nu - 3} \propto \tau^{(-\nu - 3)/4} \approx \tau^{\Gamma}.
\ee

Most known separation distributions (observed and theoretical) having $\nu \approx 1$, and therefore, for mergers driven by gravitational radiation, we generically expect $\Gamma \approx 1.$ But the exact exponent, $\Gamma$, is critical to understanding the merger rate as a function of environment and cosmic time \citep{Behroozi:2014bp}, and conversely, measurements of the DTD can constrain progenitor scenarios such as is done with Type Ia supernovae \citep{Maoz2014}. 

The functional form of $P(t_m)$ determines the delay time distribution (DTD) of a set of binaries. 
Clearly, its formulation depends on the assumptions regarding i) the distribution of the compact object binaries at birth and 
ii) whether the merging takes place solely due to the emission of GWs or other mechanisms (such as interaction with other compact objects or gas) that can accelerate the merging process \citep[e.g.][]{Antoni2019}. 
This functional form is generically defined as:
\begin{equation}
 P(\tau; \Gamma,t_{\rm min}) \propto \left\{
 \begin{array}{ll}
 0,  & \tau<t_{\rm min},\\
 \tau^\Gamma, & \tau\geq t_{\rm min},
 \end{array}
 \right.
\end{equation}
where $ t_{\rm min}$ sets the minimum delay time of the population and it is often assumed that the maximum delay time is 10 Gyr comparable to the age of the universe.
$t_{\rm min}$ is associated with the minimum separation considered plausible for a merging binary class which itself depends on the assumptions about their formation pathways.
$\Gamma$ determines the slope of the power-law distribution and reflects our uncertainty with respect to the distribution of binaries' semi-major axis at birth \citep{Zheng:2007hl,Dominik:2012cw,Safarzadeh:2019dn,Safarzadeh:2019kj,Safarzadeh:2019dp,Safarzadeh2020ApJ}.
We note that it is possible that BBHs with different masses are born with different distribution in the semi-major axis, and therefore, DTD will also be a function depending on the mass of the BBH's component. 
However, to arrive at the respect the functional form of DTD as a function of the component masses would require detailed knowledge of the assembly of the binaries, which is not analytically traceable. 

\subsection{Merger rate of a class of binaries}
\begin{figure}
\hspace{-0.2in}
\centering
\includegraphics[width=\columnwidth]{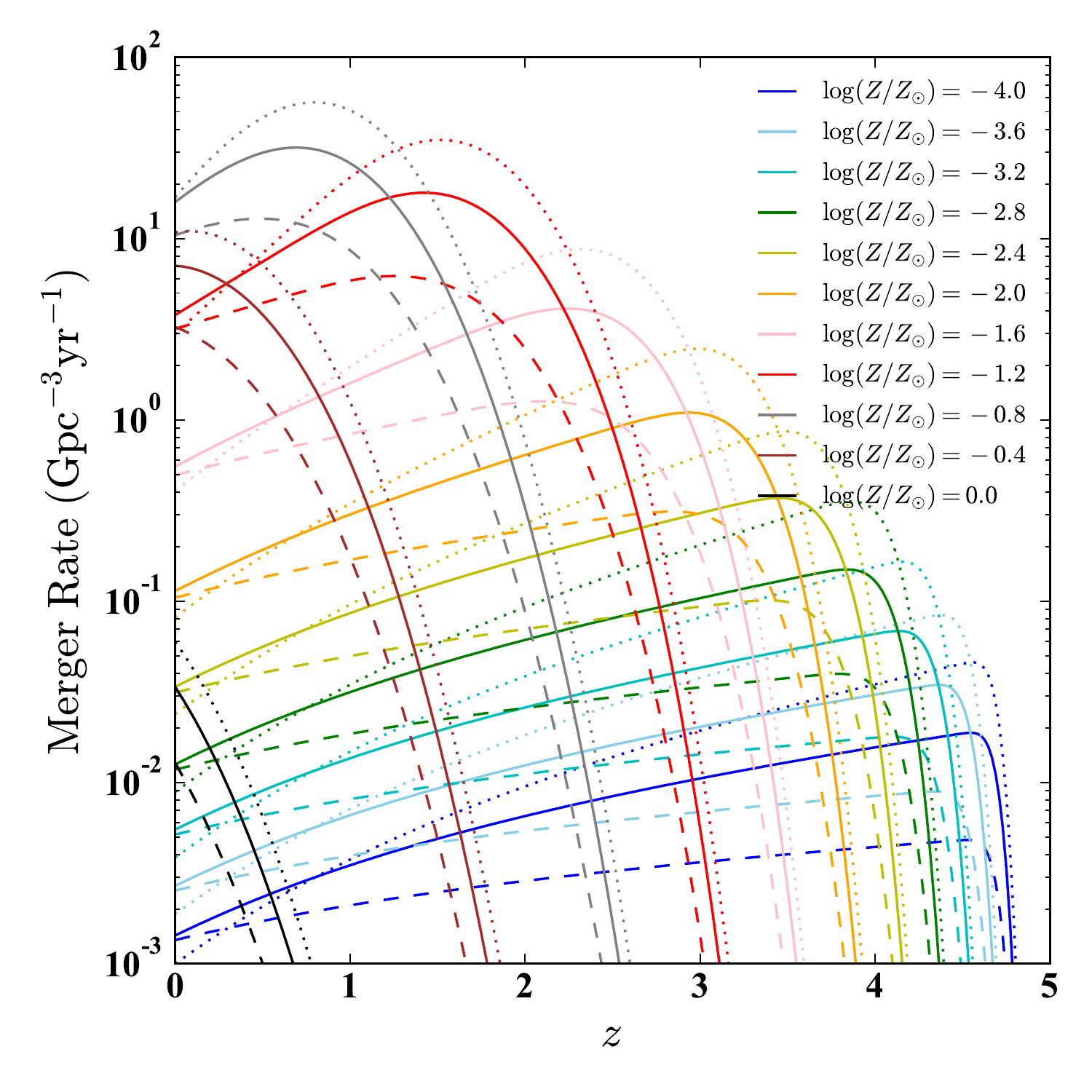}
\caption{The redshift distribution of BBH mergers formed at a given metallicity bin following a DTD parametrized with $\Gamma=-1$ and $t_{\rm min}=1~\rm Gyr$.
We assume a BBH mass efficiency of $\lambda^H_{\rm BBH}=10^{-6}$ M$_{\odot}^{-1}$. The dotted and dashed lines show the case for $\Gamma=-3/2$, and $\Gamma=-1/2$, respectively.}
\label{fig_dtd}
\end{figure}

Knowing the DTD, the expected merger rate is determined by the convolution of the respective star formation history of a class of binaries with their associated DTD:
\begin{equation}
 \mathcal{R}(\tau)=\lambda_{\rm BBH}^H\int_0^{t_{\rm max}} {\rm \psi(\tau)} P(\tau) {\rm d}\tau.
 \label{eq:rate}
\end{equation}
Here $\lambda_{\rm BBH}^H$ stands for the efficiency of the formation of BBHs that merge within a Hubble time and has the units of per solar mass. This is the same as the parameter $\eta$ defined in \citet{Santoliquido2020}.
The units of $\mathcal{R}$ is per comoving volume per time. 

\begin{align}
\mathcal{R}(z)=&\lambda_{\rm BBH}^H \int_{z_b=10}^{z_b=z} P(\tau-t_b) \psi(z_b)\frac{d\tau}{dz}(z_b)dz_b,
\end{align}
where 
\begin{equation*}
{d\tau \over dz} = - {1 \over (1+z) E(z) H_0},
\end{equation*}
\begin{equation*}
E(z)=\sqrt{{\Omega}_{m,0}(1+z)^3+{\Omega}_{k,0}(1+z)^2+{\Omega}_{\Lambda}(z)},
\end{equation*}
and $t_b$ is the time corresponding to the redshift birth $z_b$ of the CBOs.
Figure \ref{fig_dtd} shows the expected merger rate of BBHs with $\lambda_{\rm BBH}^H=10^{-6}\msun^{-1}$ born at different metallicities but all following the same DTD parametrized with $\Gamma=-1$ and $t_{\rm min}=1~\rm Gyr$.

\subsection{Mass distribution of the BBHs}
\begin{figure}
\hspace{-0.2in}
\centering
\includegraphics[width=\columnwidth]{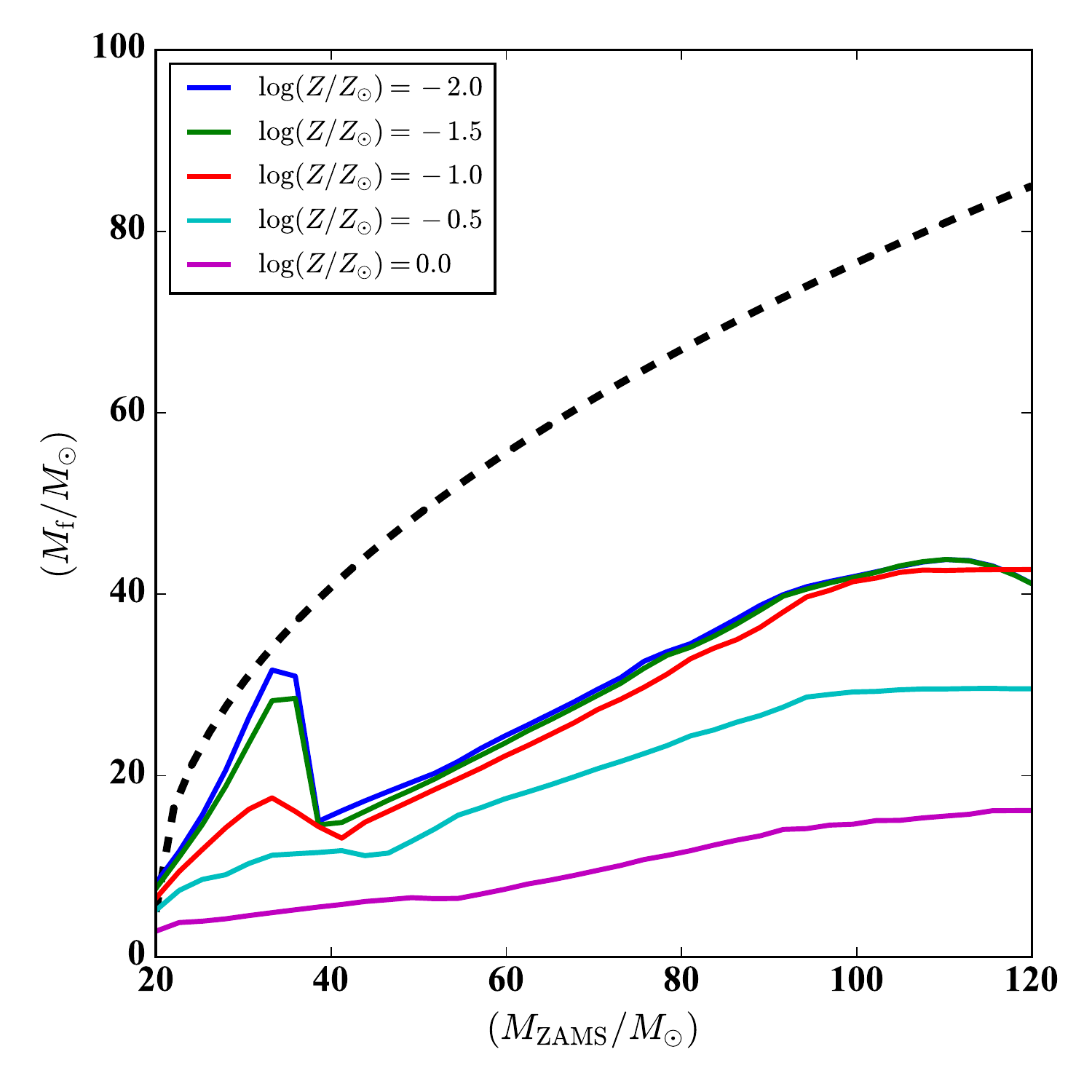}
\caption{The relation between the initial ZAMS mass of a star at a given metallicity bin and its final remnant mass adopted in this work following \citet{Zevin2020}. 
The black dashed line shows the unknown relation at $\log(Z/Z_{\odot})<-2.5$ that we will be estimating simultaneously as we fit for the BH mass function. 
At higher metallicities PPISN sets a maximum limit of about 45 $\msun$ for the final BH mass.}
\label{fig:ini_fin_relation}
\end{figure}

So far, we have only dealt with the number of BBH mergers at a given redshift formed at a given metallicity bin. But two stars sharing the same mass will form BHs with different masses if they are formed at varying metallicities. 
This is due differential wind mass loss. That is, stars born with higher metallicities lose their mass through winds more effectively, leading to the formation of a smaller mass BHs. The relation between the initial mass of the star and its final remnant is tabulated at different metallicities. To arrive at such relations, it is important to model the impact that  pair-instability supernova have on the final remnant mass. 
Solid lines in Figure \ref{fig:ini_fin_relation} show such a relation for the population synthesis code COSMIC \citep{Zevin2020}, which was derived  by adopting a the supernova model presented in \citet{Fryer2012} and imposing PPISN \citep{Woosley:2017dj}. The black line shows \emph{derived} relation at $\log(Z/Z_{\odot})=-3$ while fitting for the observed BH mass function. We such relation is similar to the results presented in \citep{Spera2015} when not modeling PPISN but at higher metallicities. 
We note that the black line does not correspond to the assumption of total collapse of a progenitor star (a one-to-one mapping between progenitor and final remnant mass before collapse).

But how are the BBHs assembled? The prevailing theory is that in a binary system, the primary (more massive) star is sampled from a Kroupa IMF and the secondary from a flat distribution. 
In this formalism, the observed mass function of BHs stems from the shape of the initial-final-mass relation and the contribution to the merger rate from different metallicity bins.

\subsection{constructing the BH mass function}
The black hole mass function in the local universe can be computed as:
\be
\frac{dN}{dm_1dtdV}=\int\frac{d\mathcal{R}(0)}{d\log(Z)} p(m_1| \log(Z))~d\log(Z),
\ee
where we use the initial-to-final mass relations convolved with the Kroupa IMF to determine $p(m_1| \log(Z))$ at each metallicity bin. In this fashion  $\lambda^H_{\rm BBH}$ also depends on metallicity.

Informed by the results of binary stellar evolution, we assume that the $\lambda^H_{\rm BBH}$ is a decreasing function of metallicity, in that formation efficiency drops at larger metallicities. 
We parametrize this relation as:
\be
\lambda^H_{\rm BBH}=a_e {\rm erf}(\log(Z)-b_e)+c_e,
\ee
with coefficients $(a_e, b_e, c_e)$ that we are going to fit for. We note that this is the formation efficiency of the BBHs that merge within a Hubble time (14 Gyrs) and not the formation efficiency of all BBHs. 
The relation between the two basically indicates what fraction of all the BBHs merges within a Hubble time. A simple estimate based on the delay time distribution of the BBHs can be that only $\approx$ 10\% do \citep{Dominik:2012cw}, which implies:
\be
\lambda^t_{\rm BBH}\approx 10~\lambda^H_{\rm BBH}.
\ee

\section{Relaxing the PISN at low metallicities }
Given that the BH mass function extends to masses above the upper limit predicted by PISN for single stellar evolution, in this work, we assume the PISN assumptions break down at sufficiently low metallicities, as argued by \citep[e.g., ][]{Costa2021MNRAS}.
Here we parametrize the relationship between the ZAMS and final remnant mass at metallicity $\log(Z/Z_{\odot})=-3$ with the following functional form:
\be
M_{\rm rm}=5+D (20-M_{\rm ZAMS})^E,
\ee
with a condition that $M_{\rm rm}$ at $M_{\rm ZAMS}=120$ is between 50 and 120 $\msun$. We call this line the Z3 line. Given this relationship, the initial-final mass is interpolated between the solid blue line shown in Figure 3 and the Z3 line for metallicities in between, 
and we assume the Z3 line holds for metallicities below $\log(Z/Z_{\odot})=-3$ as well. Therefore in this formalism we have five free parameters in our model to fit. They are  $D$, and $E$, and the three coefficients that describe the formation efficiency per solar mass as a function of metallicity ($a_e$, $b_e$, and $c_e$). 
 
\begin{figure*}
\hspace{-0.2in}
\centering
\includegraphics[width=\columnwidth]{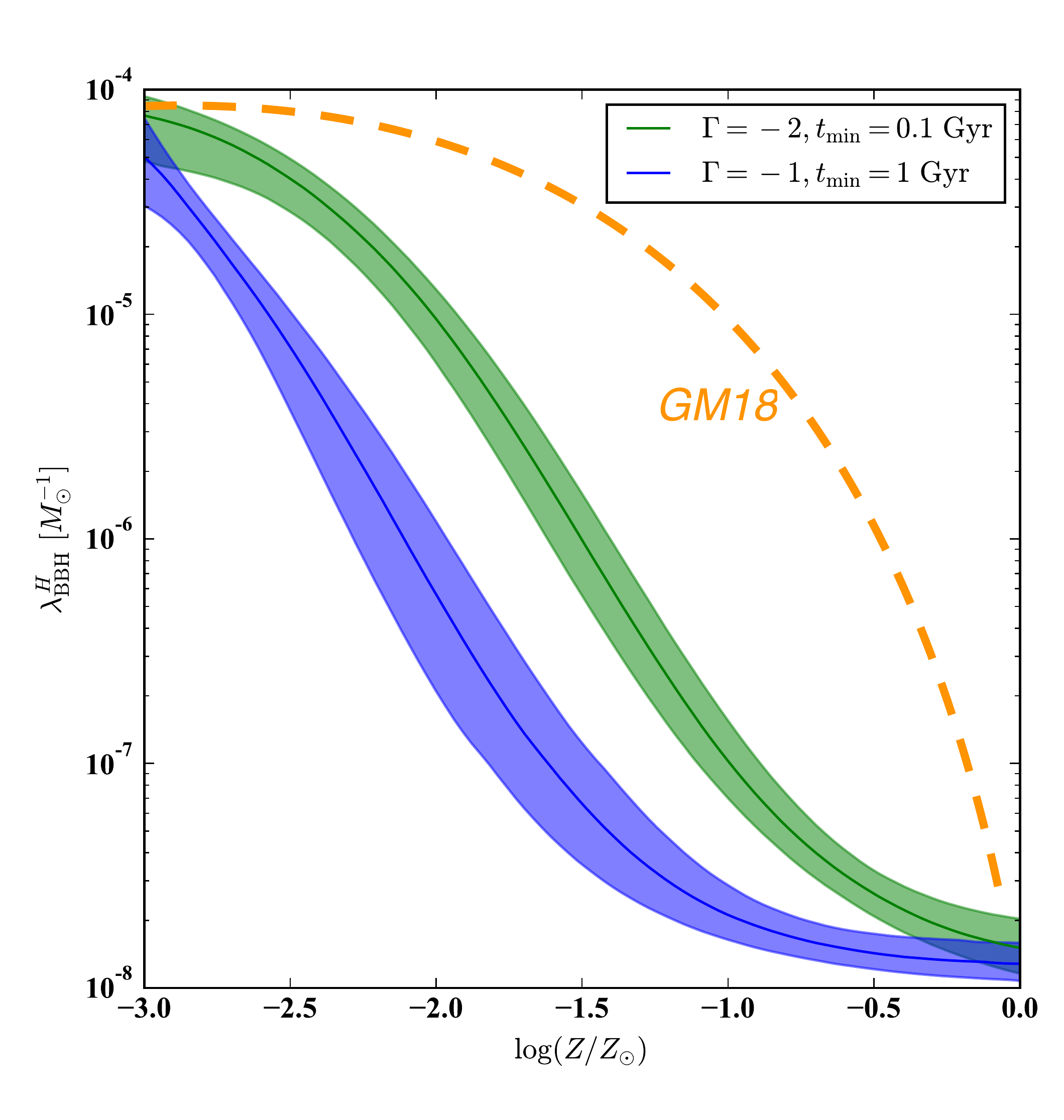}
\includegraphics[width=\columnwidth]{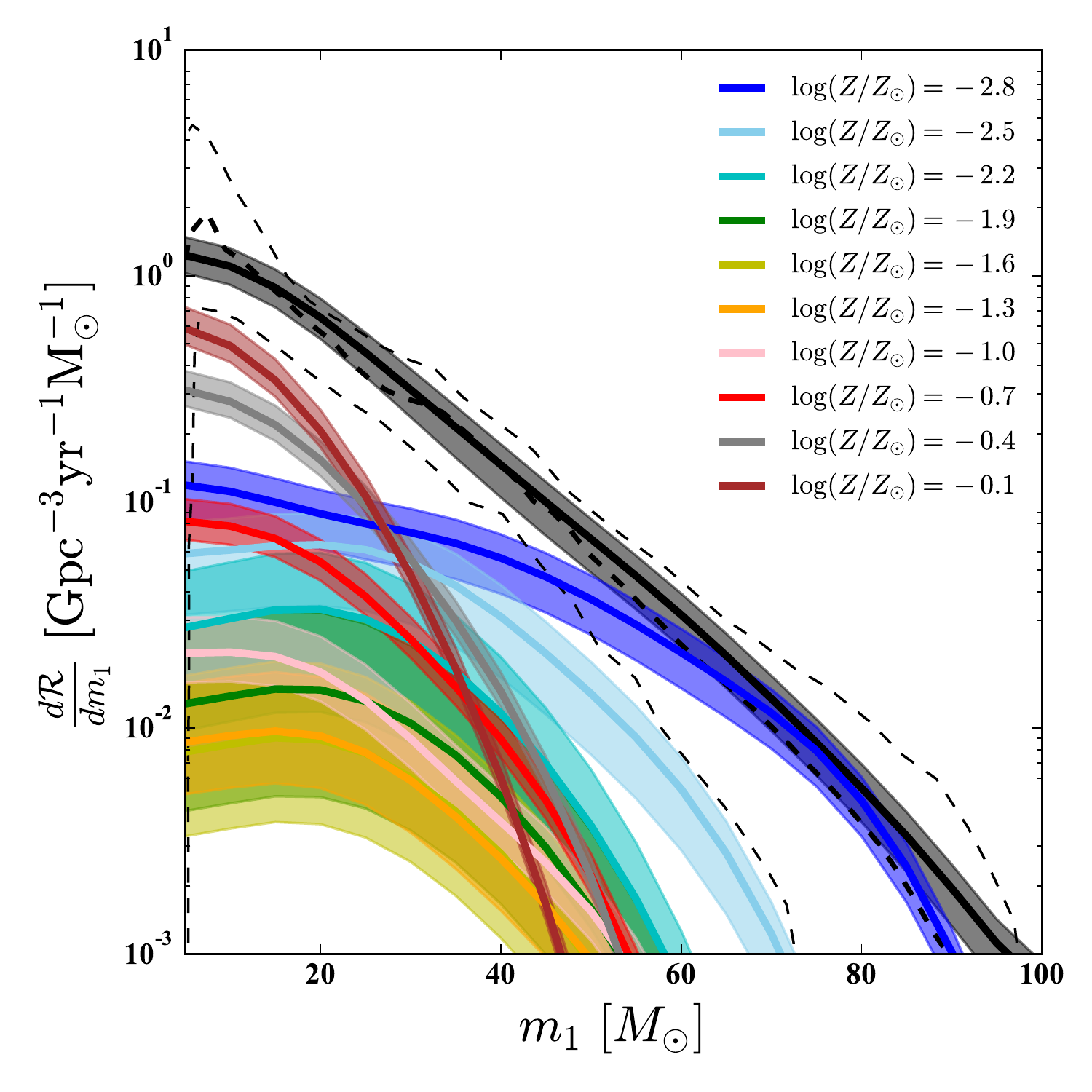}
\caption{\emph{Left Panel:} The efficiency of formation of BBHs that merge within a Hubble time for two different assumption regarding the underlying DTD. We overplot the trend found in population synthesis models \citep[][; GM18]{GM2018} with dashed orange line. \emph{Right panel:} The reconstructed BH mass function decomposed into the contributions from different metallicity bins. The black line and the corresponding shaded region indicate the mean and 16th-84th percentile of the total BH mass function. The dashed lines indicate the observed O3a BH mass function. }
\label{fig_result_1}
\end{figure*}

 Assuming two different DTDs, one with ($\Gamma=-1, t_{\rm min}=1\rm Gyr$), and the other with ($\Gamma=-2, t_{\rm min}=0.1 \rm Gyr$) we fit the observed BH mass function released in O3a rate catalog to find out the functional form of $\lambda_{\rm BBH}^H$ as a function of metallicity. We impose a prior on $\lambda^H_{\rm BBH}$ between $10^{-4}-10^{-8} M_{\odot}^{-1}$ in this study. We have tested that our results are not sensitive to the functional form adopted to model the metallicity dependence of the formation efficiency. The results are shown in Figure \ref{fig_result_1}. 

The formation efficiency drops by four orders of magnitude from about $10^{-4}$ to $10^{-8}~M_{\odot}^{-1}$ going from $10^{-3}~Z_{\odot}$ to solar metallicity which is in agreement with the trend found in population synthesis models \citep{GM2018,Santoliquido2020}, however, the exact shape is different. Assuming a steeper power for the DTD (which implies shorted merging times) would result in a high efficiency for a wider range in metallicities to compensate for the drop in the contribution at lower metallicity bins. The need for such high efficiencies can be understood by looking at the right panel where the high end of the BH mass function is accounted for by the BBHs formed at such low metallicities. The total star formation rate at such metallicities is small, and therefore a significant formation efficiency is required to compensate for it. 

\begin{figure}
\hspace{-0.2in}
\centering
\includegraphics[width=\columnwidth]{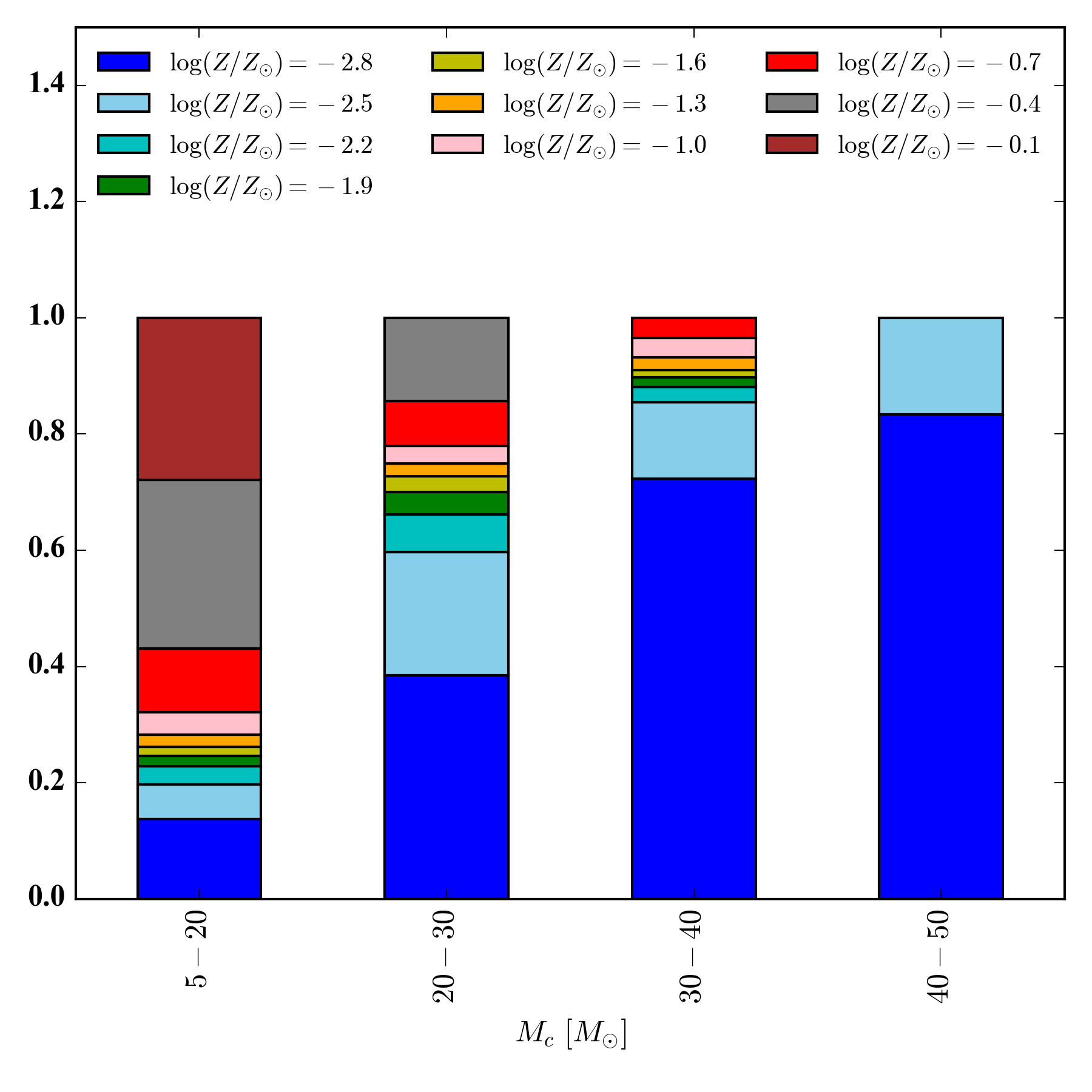}
\caption{The contribution of different metallicity bins to different chirp mass bins of the merging BBHs. BBHs with chirp masses above 40 $M_{\odot}$ are solely made from stars with metallicities less than $1/100 Z_{\odot}$. }
\label{fig_result_2}
\end{figure}

Figure \ref{fig_result_2} shows the contribution of different metallicity bins to the overall BBH statistics at different chirp masses. We have assumed the secondary BHs are drawn from ZAMS stars with a flat distribution. The contribution from low metallicity stars starts to dominate at larger chirp masses such that BBHs with chirp masses above about 40 $M_{\odot}$ are solely made from stars with metallicities below $1/100 Z_{\odot}$. We note that this is an imprint of the fact that BBHs with such high masses need to be assembled from low metallicity stars due to the adopted initial-to-final mass relation.

\begin{figure}
\hspace{-0.2in}
\centering
\includegraphics[width=\columnwidth]{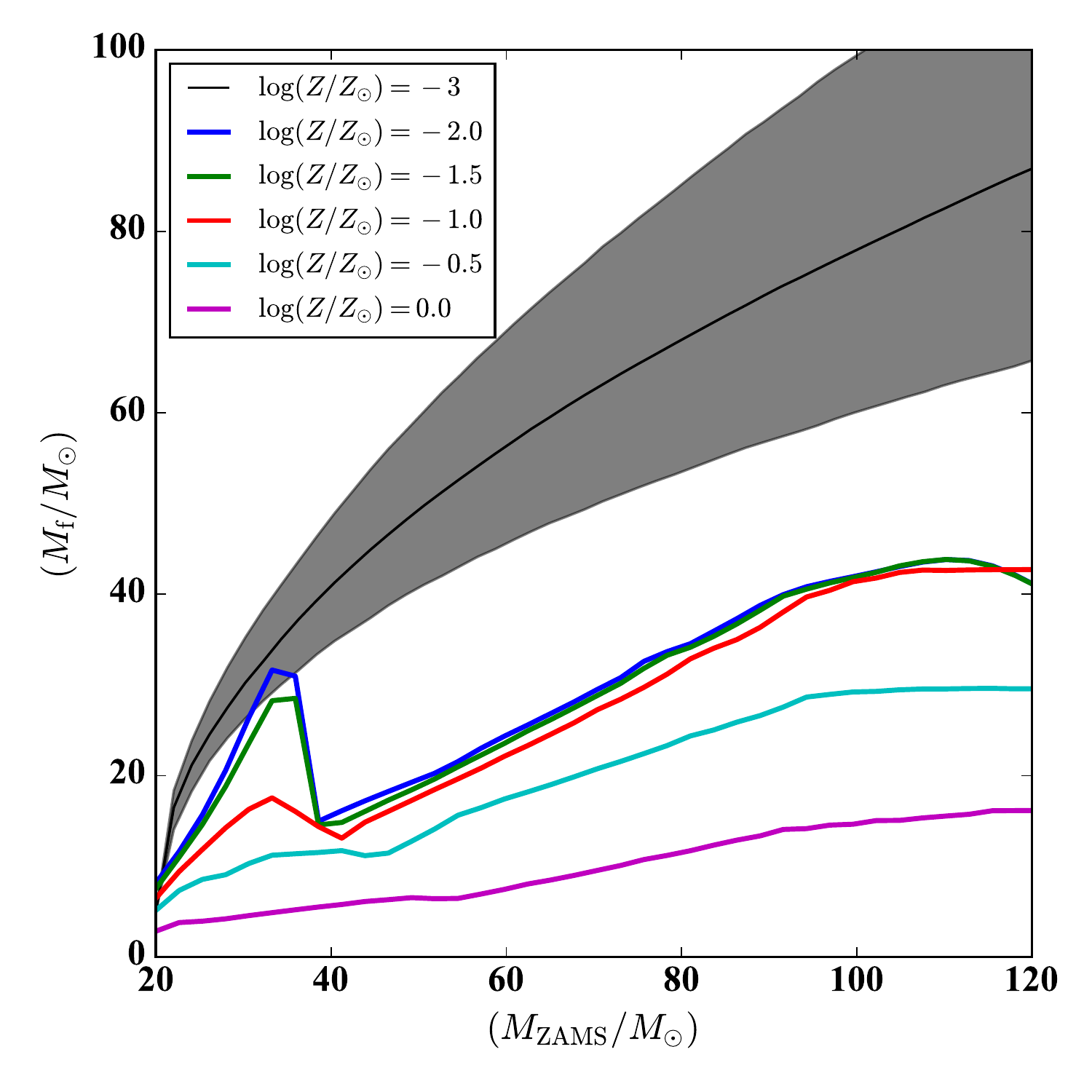}
\caption{The posterior Z3 line relation from our fits to the BH mass function shown with the black line and its 16th-84th percentile with the shaded region. Other lines are similar to what is shown in Figure 3.}
\label{fig:Z3line}
\end{figure}

Figure \ref{fig:Z3line} shows posterior for the Z3 line relation is shown with black line and the shaded region indicates the 16th-84th percentile. 
A steeper line would form too massive BHs inconsistent with the observed BH mass function. 
While this picture can make sense, we have to ask whether such inferred high BBH formation efficiencies are viable with a universal (metallicity independent) Kroupa IMF.

Assuming the stars are formed according to a universal Kroupa IMF ($\kappa(M)$), the fraction of stars born massive enough to from a BH at the end of their life ($M\geq20\msun$) is given by:
\be
f^m_{BH}=\frac{\int_{M_{\rm min}}^{M_{\rm max}} \kappa(M) dM}{\int_{0.01\msun}^{M_{\rm max}} \kappa(M) dM}.
\ee
Adopting a Kroupa IMF, this integral amount to $f^m_{BH}\approx10^{-3}$, assuming $M_{\rm min}=20~\msun$ and $M_{\rm max}=150~\msun$ for BH progenitors. 
The average stellar mass in Kroupa IMF is given by:
\be
\bar{M_*}=\frac{\int_{0.01\msun}^{150\msun} \kappa(M) M dM }{\int_{0.01\msun}^{150\msun} \kappa(M) dM}\approx0.4~M_{\odot}
\ee
Therefore, the formation efficiency of a single BH is given by:
\be
\lambda_{\rm BH}=\frac{f^m_{BH}}{\bar{M_*}}\approx3\times10^{-3}~M_{\odot}^{-1}.
\ee

Of these stars, a given fraction will be born in a binary system ($f_b$). Typical assumption is $f_b=0.5$ \citep{Sana2012}. 
These secondary stars are assumed to have a flat IMF, meaning their average mass is half of the primary star's mass in the binary. The average of the primaries can be calculated as:
\be
\bar{M^{p}_*}=\frac{\int_{20\msun}^{150\msun} \kappa(M) M dM }{\int_{20\msun}^{150\msun} \kappa(M) dM}\approx40~M_{\odot}.
\ee
As such, the formation efficiency of a BBH system can be written as:
\be
\lambda_{\rm BBH}=f_b \frac{f^m_{BH}}{\bar{M_*}+f^m_{BH}\frac{\bar{M^{p}_*}}{2}}\approx10^{-3}~M_{\odot}^{-1}.
\ee

We note that this is the $\emph{total}$ BBH formation efficiency, out of which about 10\% are expected to be born such that they merge within a Hubble time \citep[e.g., ][]{Dominik:2012cw}.
Therefore the formation efficiency of the BBHs merging within a Hubble time is ten times smaller:
\be
\lambda_{\rm BBH}^H\approx10^{-4}~M_{\odot}^{-1}.
\ee
We arrive at this number without considering possible imparted kicks to the BHs, which, if it is the case, will further reduce the formation efficiency.

What we obtain by fitting the BH mass function is $\lambda_{\rm BBH}^H\approx 10^{-4}~M_{\odot}^{-1}$ which is close to this limit. 
It is thus possible that considering a flatter IMF at lower metallicities can help widen the gap between the maximum theoretical limit and what it takes to fit the observed BH mass function.
Such flat IMFs have been observed to be the case from numerical simulations of star formation at low metallicities \citep[e.g., ][]{Hirano+2015} than the IMF observed in the local universe \citep{Salpeter1955}. 
This is expected due to the fact that lack of metals suppresses fragmentation of the star-forming gas, leading to the formation of the massive stars \citep[e.g.][]{Nakamura2001,Clark2011}.

\section{Results with Gas accretion at low metallicities}

In the above discussion we have assumed that low metallicity stars ($\log(Z/Z_{\odot})<-2.5$) can avoid pair-instability SNe and, therefore, are capable of forming BHs with masses above 50 solar masses. 
However, how could we account for the observed merger rate at large BH masses if we assume all  stars irrespective of their birth metallicity  have to go through the PISNe? One option would be to accrete gas from the ambient environment such that the BBH doubles its mass before merging. But what fraction of the BBHs would experience such accretion phases?

\begin{figure*}
\hspace{-0.2in}
\centering
\includegraphics[width=\columnwidth]{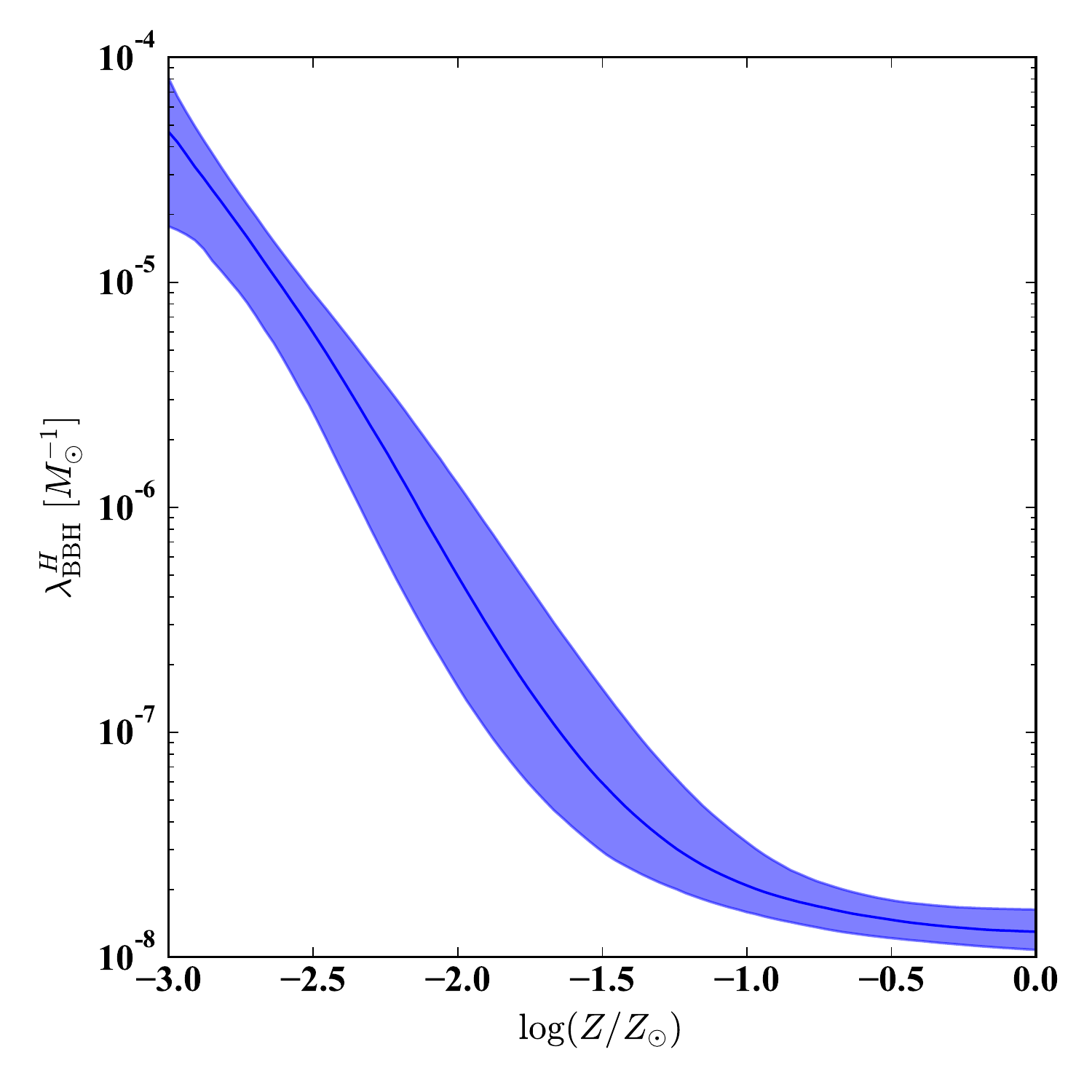}
\includegraphics[width=\columnwidth]{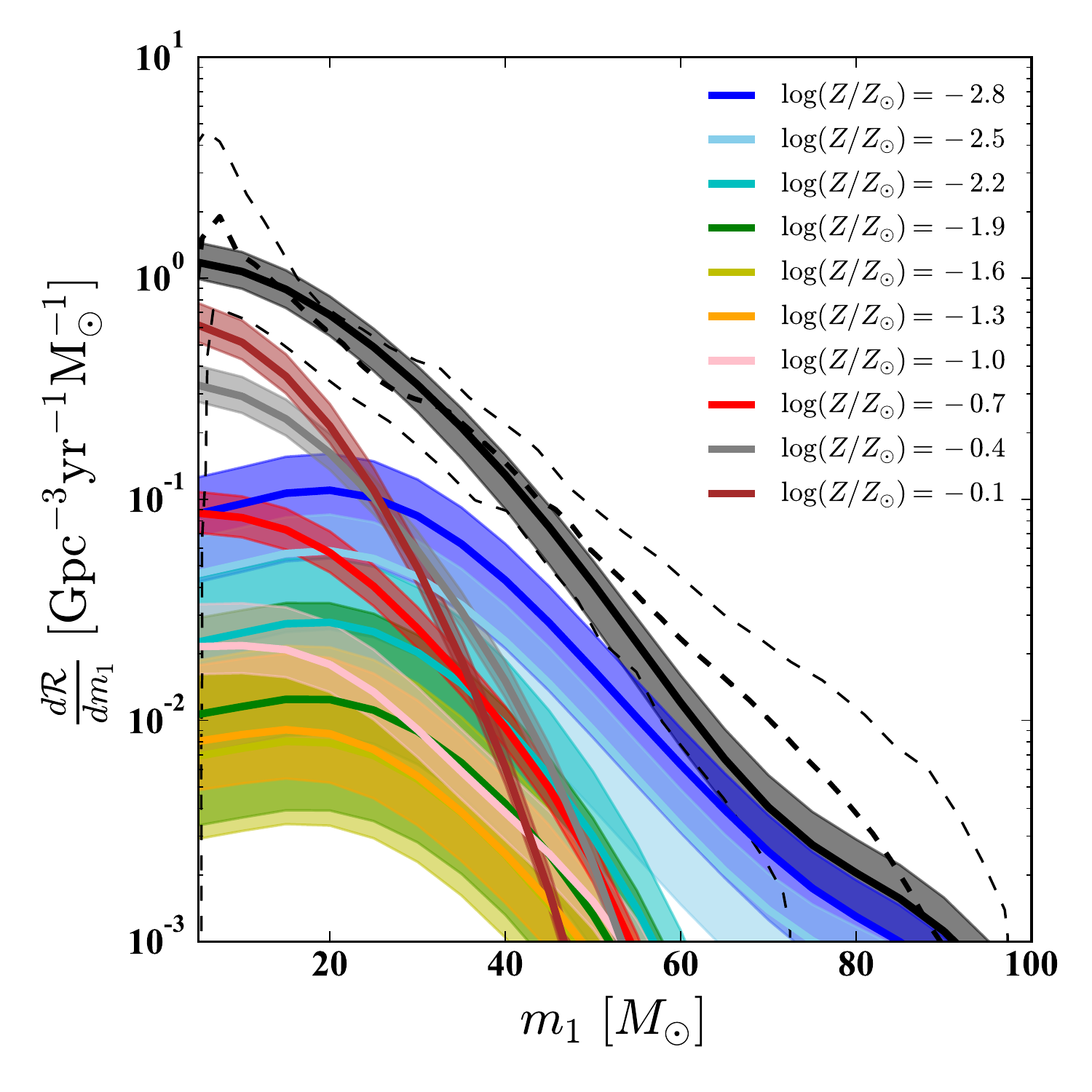}
\caption{\emph{Left Panel:} The efficiency of formation of BBHs that merge within a Hubble time assuming PISNe and a universal Kroupa IMF and considering 10\% of the BBHs with $\log(Z/Z_{\odot})<-2.5$ to undergo a phase of mass accretion doubling their mass prior to merging event \emph{Right panel:} The reconstructed BH mass function decomposed into the contributions from different metallicity bins. The black line and the corresponding shaded region indicate the mean and 16th-84th percentile of the total BH mass function. The dashed lines indicate the observed O3a BH mass function. }
\label{fig_result_3}
\end{figure*}

\begin{figure}
\hspace{-0.2in}
\centering
\includegraphics[width=\columnwidth]{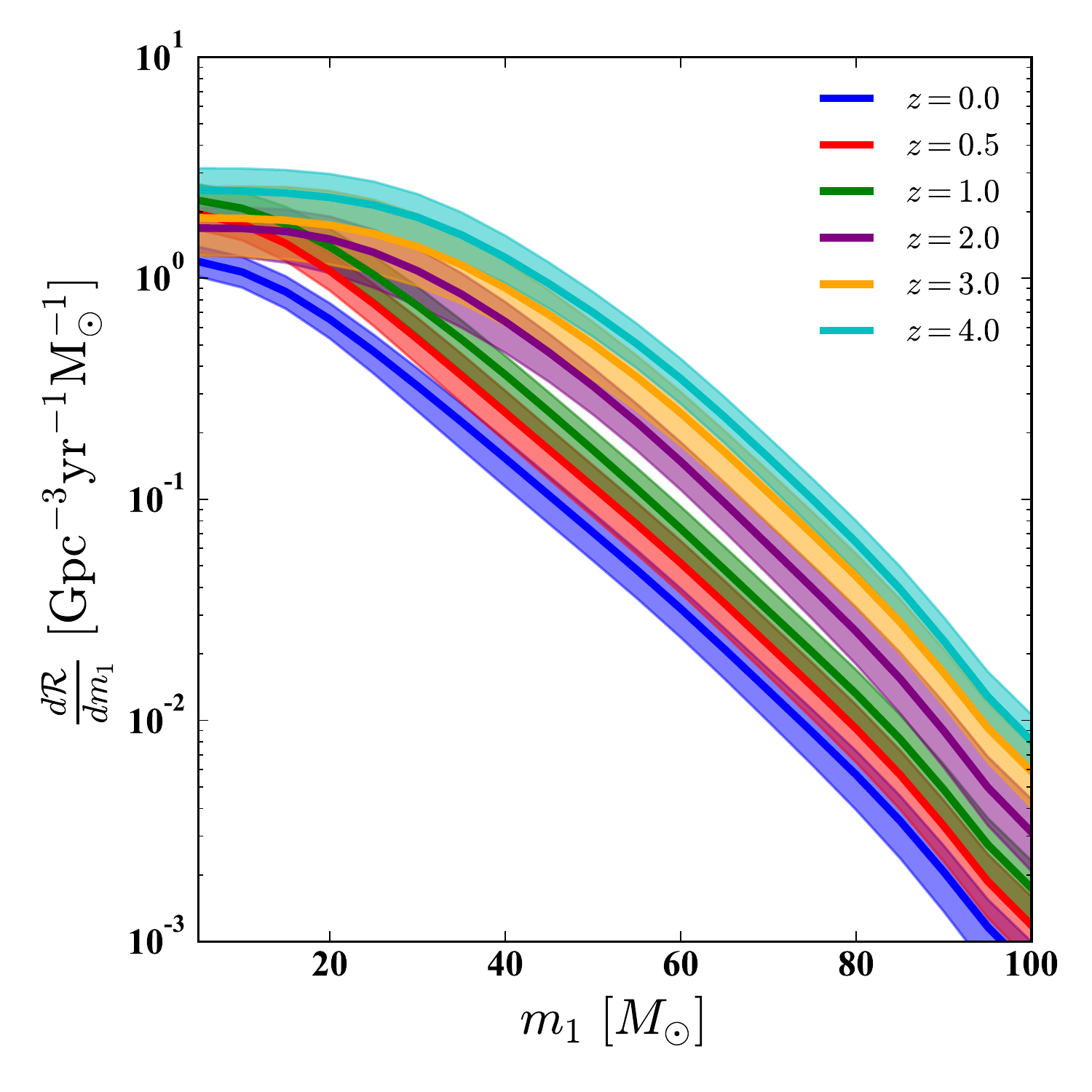}
\caption{The predicted redshift evolution of the mass function which is similar whether we relax PPISN at low metallicities or impose it but assuming 10\% of the BBHs at low metallicities double their mass before merging due to ambient gas accretion in dark matter halos at high redshifts.}
\label{fig_compare}
\end{figure}

\begin{figure}
\hspace{-0.2in}
\centering
\includegraphics[width=\columnwidth]{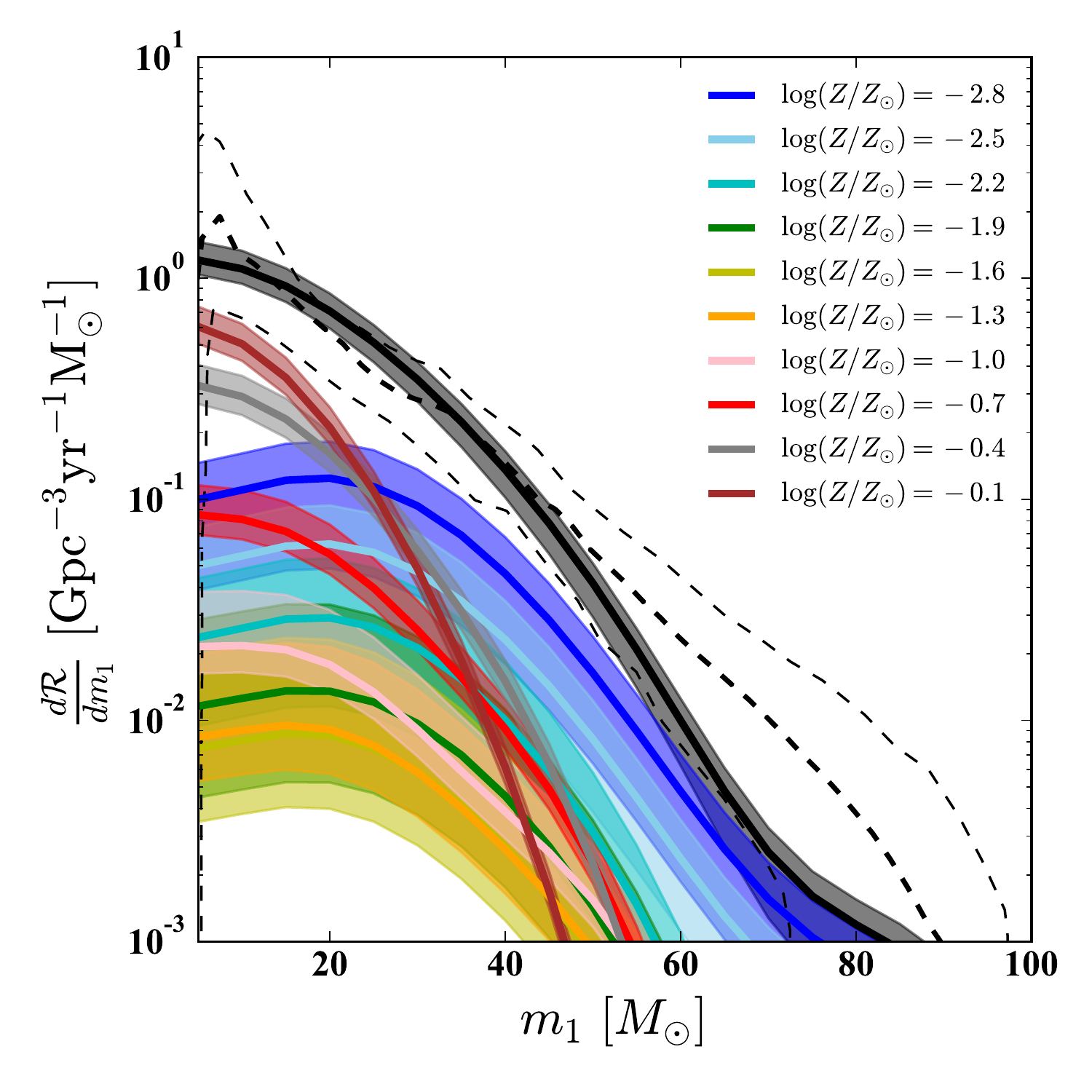}
\includegraphics[width=\columnwidth]{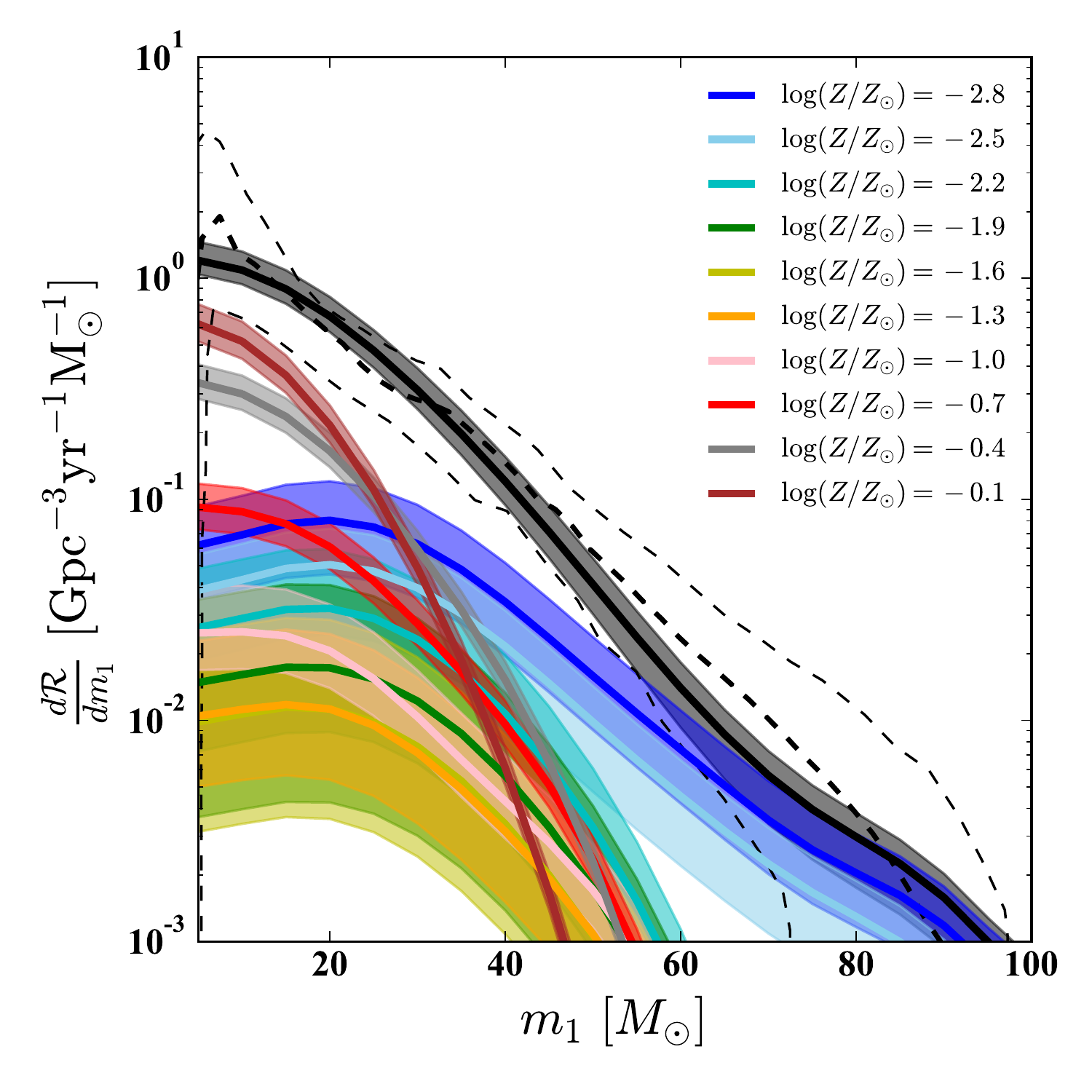}
\caption{\emph{Top Panel:} The reconstructed BH mass function decomposed into the contributions from different metallicity bins assuming $f_a=5\%$. The black line and the corresponding shaded region indicate the mean and 16th-84th percentile of the total BH mass function. The dashed lines indicate the observed O3a BH mass function. \emph{Bottom panel:} showing the same for $f_a=20\%$. }
\label{fig_result_x}
\end{figure}

\begin{figure*}
\hspace{-0.2in}
\centering
\includegraphics[width=0.65\columnwidth]{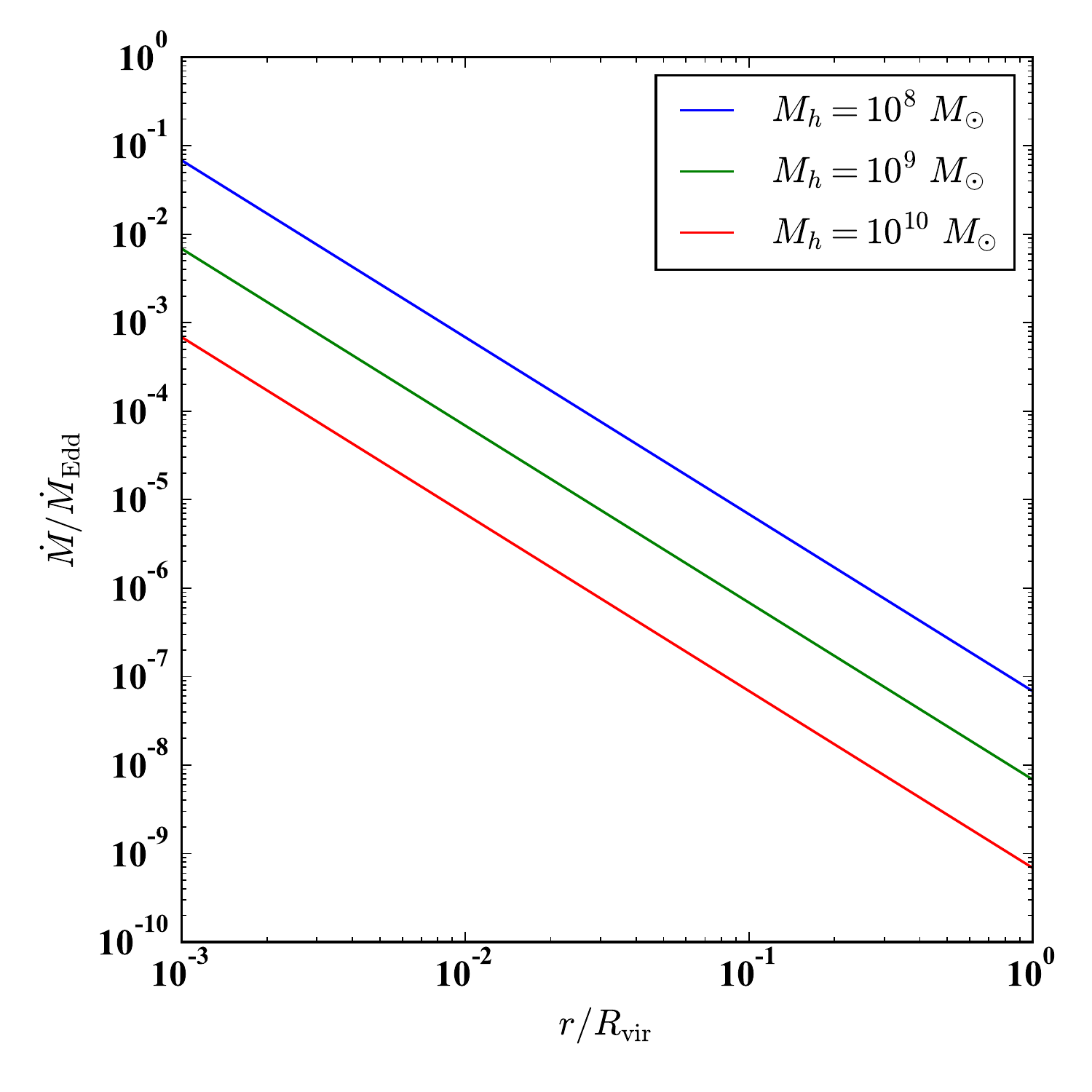}
\includegraphics[width=.65\columnwidth]{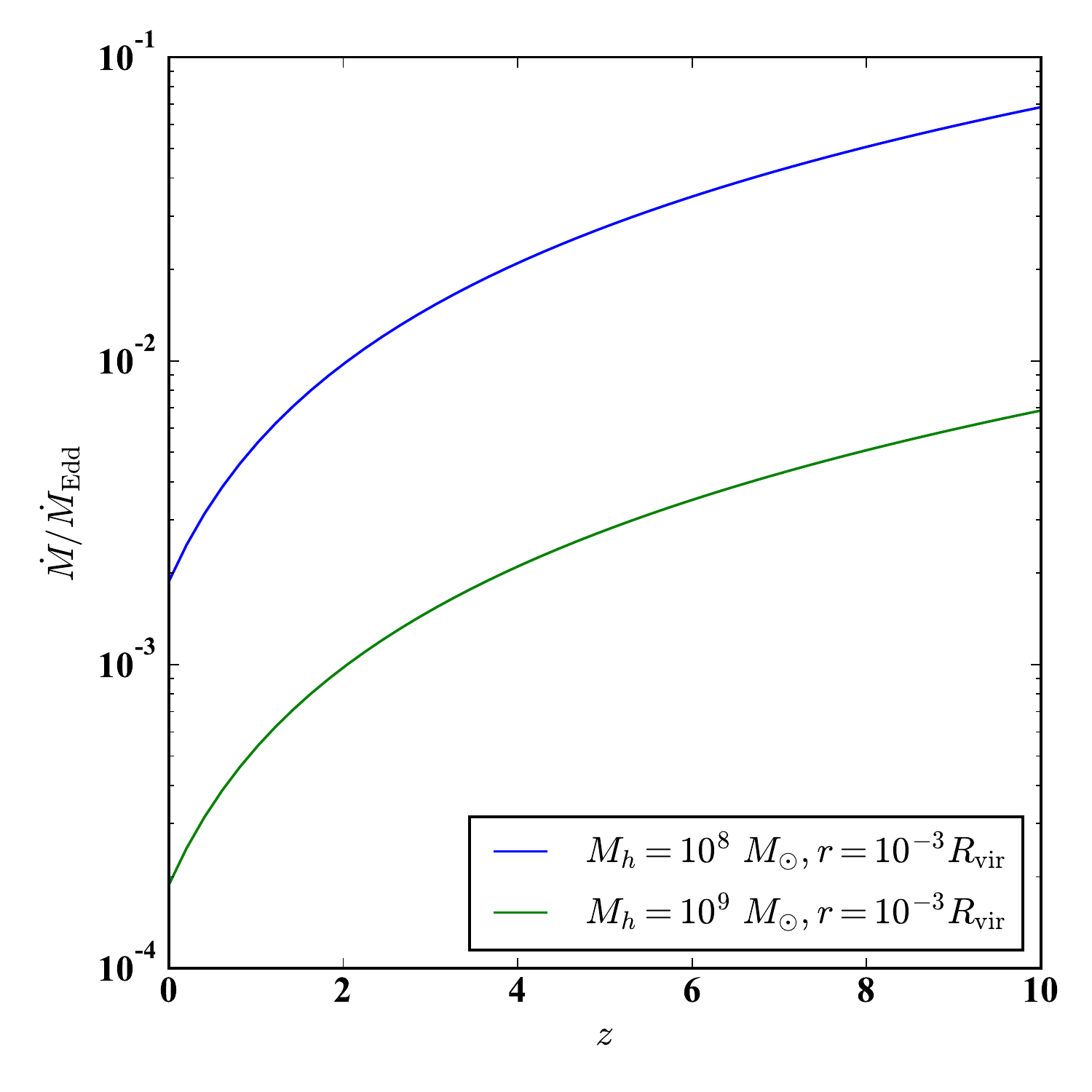}
\includegraphics[width=.65\columnwidth]{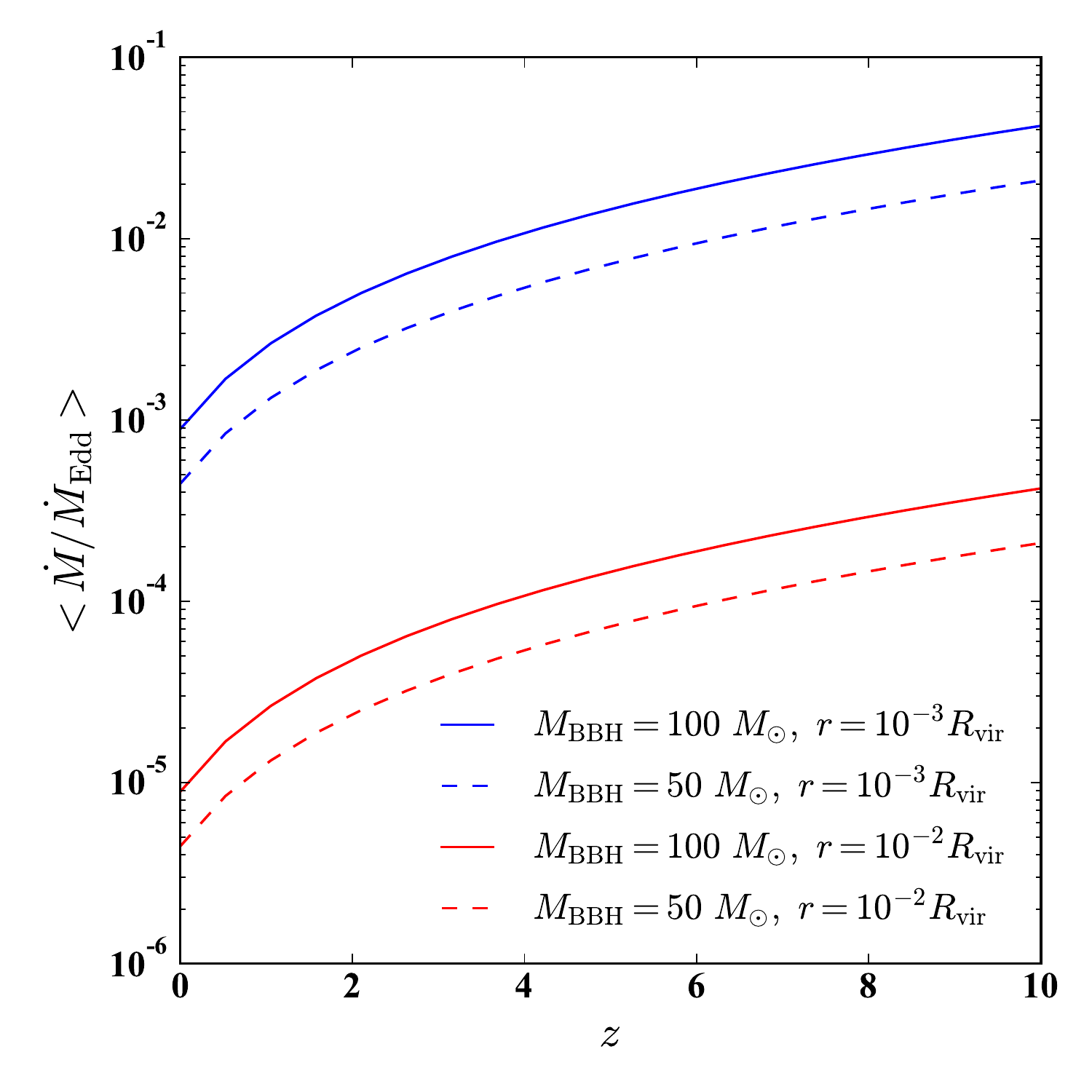}
\caption{{\emph Left panel:} The ratio of BHL accretion rate onto a BBH of total mass 100 $\msun$ as a function of its distance (normalized to the virial radius of its host dark matter halo) from the center of the dark matter halo for different halo masses at redshift $z=10$. {\emph Middle panel:} The ratio of BHL accretion rate onto a BBH of total mass 100 $\msun$ at a fixed distance (normalized to the virial radius of its host dark matter halo) from the center of the dark matter halo for different halo masses.
 {\emph Right panel:} The halo mass function weighted ratio of the accretion rate onto a BBH of a given mass over its Eddington limit as a function of redshift. The ratio of BHL accretion rate onto a BBH of total mass 100 $\msun$ at a fixed distance (normalized to the virial radius of its host dark matter halo) from the center of the dark matter halo for different halo masses.}
\label{fig_accr_explore}
\end{figure*}

In order to study this we parametrize  the fraction $f_a$ of BHs at each metallicity bin  that they are able to double their mass via accretion. Figure \ref{fig_result_3} shows similar results to Figure \ref{fig_result_1} but assuming $f_a=0.1$ (10\% of all the BBHs with $\log(Z/Z_{\odot})<-2.5$) undergo a phase of accretion that doubles their mass prior to merging. 
In this scenario, the formation efficiency of BBHs merging within a Hubble time falls within the range compatible with a metallicity independent Kroupa IMF at all metallicities. 

We note that we have assumed a constant fraction of 10\% of the BBHs double their mass prior to merging if born at very low metallicities. BBHs formed in the early universe have more time to merge by redshift zero, and therefore, a larger fraction of such BBHs can undergo a phase of mass accretion compared to the BBHs formed at higher metallicities \citep[e.g., ][]{SH2020}. Without detailed modeling of gas accretion's impact on the mass distribution of the merging BBHs, our simple approach indicates that mass accretion, though impacting only a small subset of the BBHs, can change the progenitor metallicity distribution of the merging BBHs. We defer a more detailed analysis of this type (such as a metallicity-dependent gas accretion scenario) to future work. Figure \ref{fig_compare} shows the predicted mass function from either of the two models we explored in this work.  

Figure \ref{fig_result_x} shows the impact of assuming $f_a=5\%$ (top panel), and $f_a=20\%$ (bottom panel) for the reconstructed BH mass function. The results indicate that we at least need 10\% of the BHs at low metallicities to double their mass, while a larger fraction can not be ruled out. Is a fraction of $f_a=10\%$ a reasonable assumption? 
The framework relies on the computation of the Bondi–Hoyle–Lyttleton \citep[BHL, ][]{Bondi:1952fc,Bondi:1944gc,Hoyle:1939fl} rate on a single BBH which is given by:
\be
\dot{M}_{\rm BBH}=\frac{4\pi G^2 M_{\rm BBH}^2 \rho_{g}}{(c_s^2+v_{\rm rel}^2)^{3/2}},
\ee
where $c_s$ is the sound speed in the medium and $v_{\rm rel}$ is the relative speed of the BH and the surrounding medium. Throughout this work, we treat the BBH as a single point mass. 
We assume the BBHs are born in dark matter halos, with isothermal gas density profile given by:
\be
\rho_{g}(r)=f_b\frac{V_{\rm vir}^2}{2 \pi G r^2}
\ee
where $f_b=\Omega_b/\Omega_{DM}=0.16$ is the cosmological baryon fraction, and $V_{\rm vir}$ is the virial velocity of the dark matter halo given by:
\be
V_{\rm vir}=23 \left(\frac{M}{10^8\msun}\right)^{1/3} \left(\frac{1+z}{10}\right)^{1/2} h^{1/3} \kms. 
\ee
The Virial temperature of the halo is given by:
\be
T_{\rm vir}=1.98\times 10^4 \frac{\mu}{0.6} \left(\frac{M}{10^8\msun}\right)^{2/3} \left(\frac{1+z}{10}\right) h^{2/3} K,
\ee
and the sound speed is related to $T_{\rm vir}$ as $c_s=\sqrt{3 K_BT_{\rm vir}/(\mu m_p)}$, where $m_p$ is proton's mass and for a primordial gas $\mu=1.2$.

We construct a halo mass function \citep{ST1999} by:
\be
dn/dM=\sqrt{\frac{2}{\pi}} \frac{\rho_m}{M} \frac{-d ln(\sigma)}{dM} \nu_c e^{-\nu_c^2/2},
\ee
where $\nu_c=\delta_{\rm crit}(z)/\sigma(M)$. $\delta_{\rm crit}(z)$ is the critical collapse density at redshift $z$, and $\rho_m$ is the comoving matter density. $\sigma(M)$ is given by:
\be 
\sigma^2(M)= \sigma^2(R)= \int_0^{\infty}\frac{dk}{2 \pi^2} \,k^2 P(k) w^2(kR)\ ,\label{eqsigM}
\ee 
where $M = 4 \pi R^3 \Omega_M \rho_c/3$, $w(kR)\equiv 3 j_1(kR)/(kR),$ with $j_1(x)\equiv(\sin x-x\cos x)/x^2,$, and $\rho_c=3H_0^2/8\pi G$ is the critical density of the universe. $D(z)$ is the linear growth factor:
\be
D(z)\equiv\frac{H(z)}{H(0)} \int_z^{\infty} \frac{dz^\prime (1+z^\prime)}{H^3(z^\prime)} \Bigg [\int_0^{\infty} \frac{dz^\prime (1+z^\prime)}{H^3(z^\prime)}\Bigg]^{-1},
\ee
and $\delta_{\rm crit}(z)=1.686/D(z)$. The above scaling relations are adopted following \citet{LoebFurlanetto013}.

To build intuition, we first compute the fraction of BHL accretion rate over the Eddington rate of a BBH which by assuming 10\% radiative efficiency is given by:
\be
\dot{M}_{\rm Edd}=2.2\times10^{-8} \left(\frac{M}{M_{\odot}}\right) M_{\odot} \rm yr^{-1},
\ee
Left panel of figure \ref{fig_accr_explore} shows the ratio of the BHL accretion over Eddington limit for a 100 $\msun$ BBH located in dark matter halos of different mass at redshift $z=10$. 
The location is normalized to the virial radius of the host dark matter halo. Alternatively, we can look at the redshift evolution of the accretion ratio at a fixed normalized distance from the galactic center at a fixed dark matter halo mass.
This is show in the middle panel of Figure \ref{fig_accr_explore}. Next step would be to compute the weighted accretion ratio given the halo mass function, by defining $\psi_{\rm BBH}=\dot{M}_{\rm BBH}/\dot{M}_{\rm Edd}$:
\be
<\psi_{\rm BBH}(z)>=\frac{\int dn/dM(z)~\psi_{\rm BBH}(M)~dM}{\int dn/dM(z)~dM},
\ee
which is computed at a fixed BBH mass and a given distance from the center of the dark matter halo. 

The result is shown in the right panel of Figure \ref{fig_accr_explore} where a 100 $\msun$ BBH is effectively accreting at a few percent of its Eddington limit if located within $10^{-3}~R_{\rm vir}$ of its host halo at redshifts above $z>5$. The results are sensitive to the distance of the BBH from the center of the dark matter halo as a ten-fold increase in the distance leads to about two orders of magnitude drop in the accretion rate. 
Assuming an accretion rate of $\dot{M}\approx 0.03 \dot{M}_{\rm Edd}$ for a BBH, it takes about 1 Gyr for the binary to double its mass, and within a Gyr timescale, a BBH can increase its mass by about 20\% while accretion at 1\% of the Eddington limit.
Therefore, based on the results we have presented, we would require about 10\% of the BBHs at high redshift to be born within about one parsec from the center of their atomic cooling halos and remian in such halos for about a Gyr. Whether this is plausible remains to be explored with future high-resolution cosmological simulations of cosmic gas accretion onto atomic cooling halos where star formation at sub-parsec scale can be resolved. 

The complicated dynamic of a BBH both in terms of gas accretion and dynamical friction combined with radiative efficiency remains a challenge to be simulated and goes beyond the simple analytic calculations presented here. 
Three-dimensional simulations that have attempted to follow the trajectories of remnants BHs in early protogalaxies~\citep[e.g.][]{Alvarez+2009,Smith+2018,Pfister+2019} have found that they spend long periods in low-density regions away from the cores of the parent halos. Nevertheless, during the wild journey of the BBH in such environment, the BBH can pass through the central regions of an atomic cooling halo where only few thousands of years is enough for the BBH to double its mass.

\section{Summary and Discussion}

\begin{figure*}
\hspace{-0.2in}
\centering
\includegraphics[width=\columnwidth]{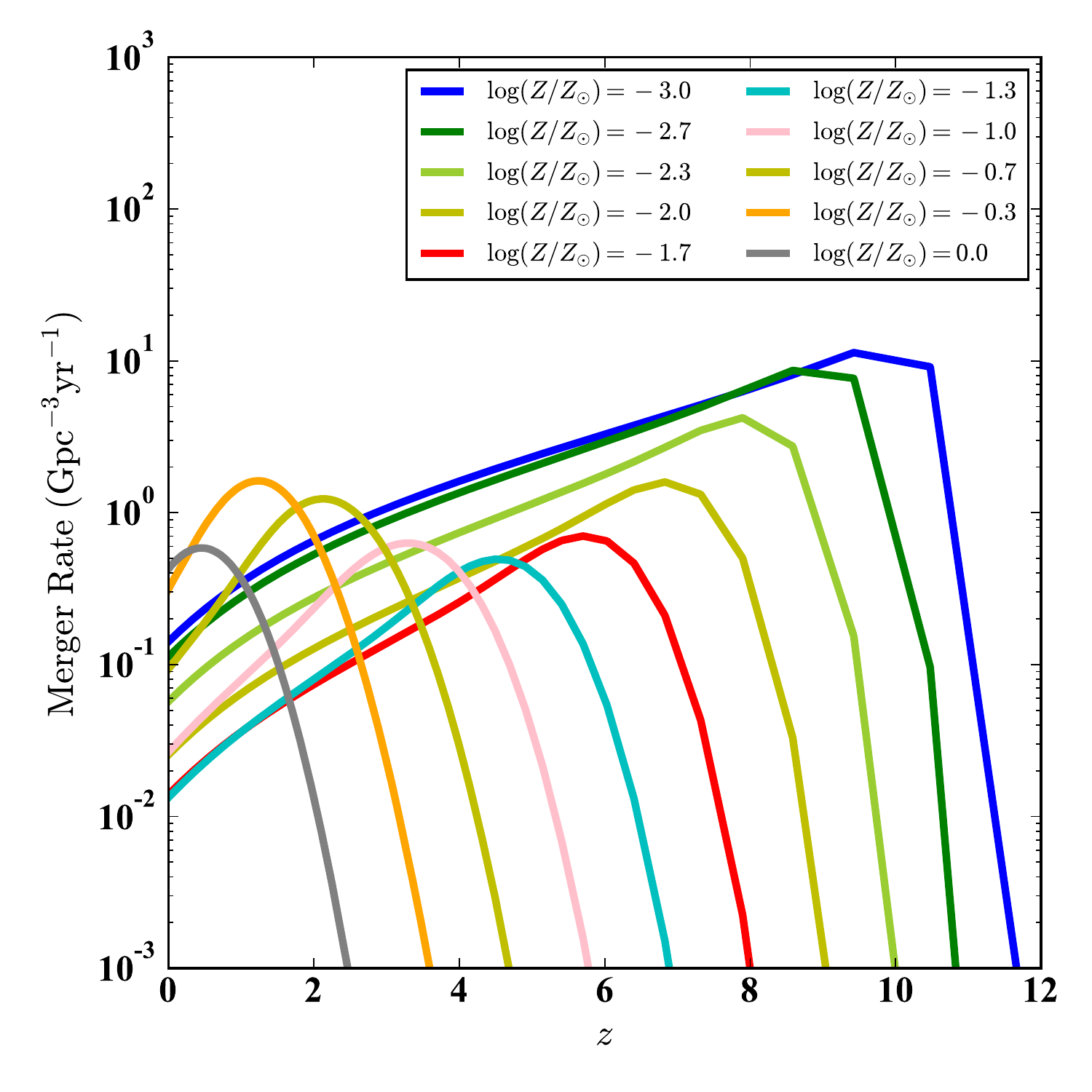}
\includegraphics[width=\columnwidth]{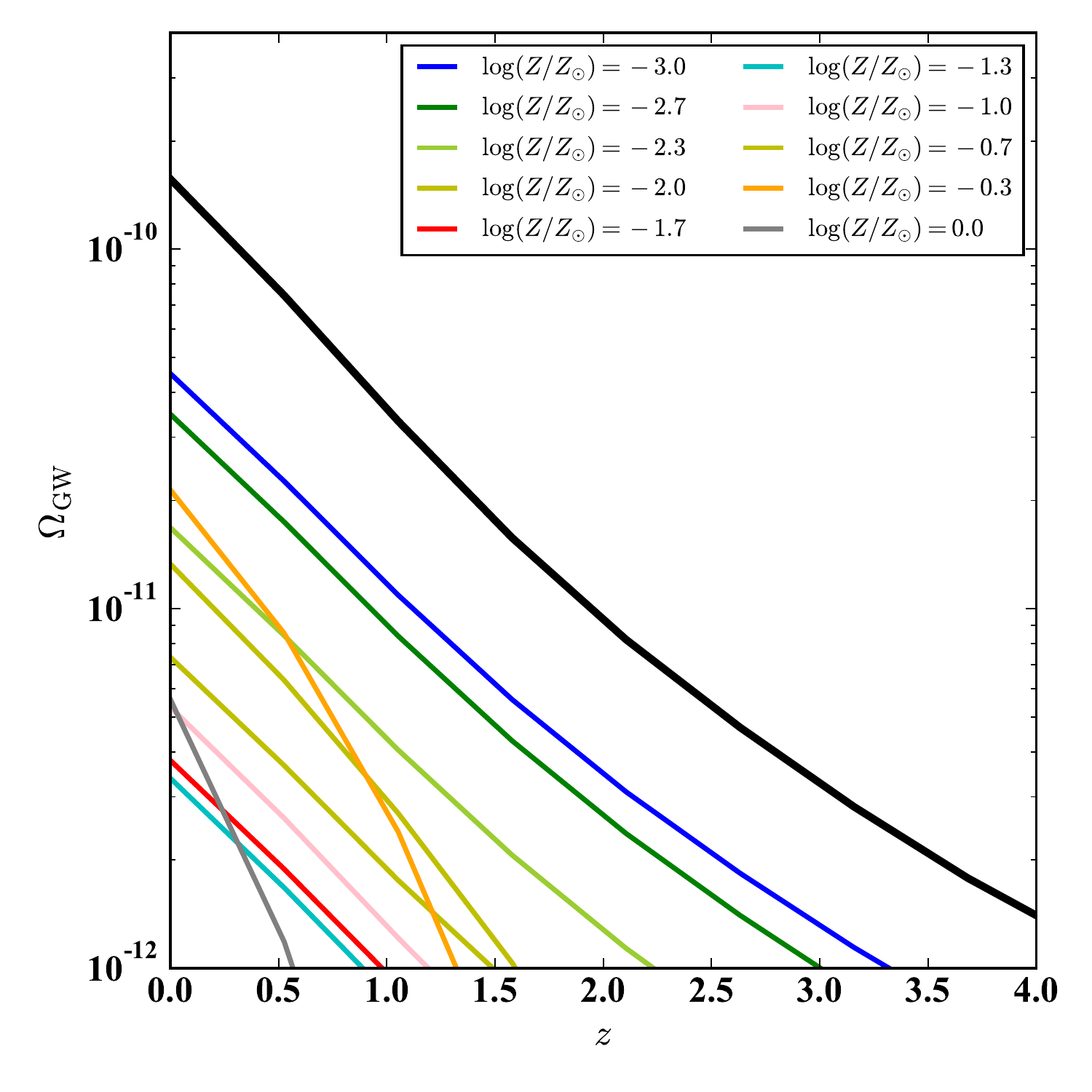}
\caption{\emph{Left Panel:} the predicted redshift evolution of the BBH merger rate assuming 10\% of the BBHs at low metallicities double their mass before merging due to ambient gas accretion in dark matter halos at high redshifts. \emph{Right panel:} the predicted redshift evolution of $\Omega_{\rm GW}$ as a function of redshift for BBHs formed in different metallicity bins and the overall which is plotted with black line. }
\label{fig_discuss}
\end{figure*}

Due to the suppression of wind mass loss, a star born at low metallicities form a more massive BH compared to if the star was born in a higher metallicity environment. While this is, in general, a correct physical trend, whether the pair-instability limit is operating at low metallicities remains a matter of dispute. If we assume there is a sharp upper mass limit of about 50 solar masses at \emph{all} metallicities we would not be able to explain the observed BH mass function solely through the binary stellar evolution as it extends to BHs with masses about 100 $M_{\odot}$. In this case, we would need to either form more massive BHs through mergers of lower mass BHs in dense environments or invoke gas accretion. While we don't investigate the dynamical formation channel, we focus on the gas accretion scenario at low metallicities as an alternative to forming massive BHs. The gas accretion is only relevant at low metallicities since the universe is much denser at higher redshifts, where the star formation is primarily taking place in low metallicity environments.

However, if we assume pair-instability can be surpassed in the lowest metallicities, meaning BHs with masses about 100 $\msun$ can form directly from the stellar collapse, then we can explain the BH mass function through binary stellar evolution. Given that such massive BHs would form only at the lowest metallicities, the observed merger rate of such massive BHs will inform us about the formation efficiency of such BHs at highest redshifts (which correspond to the lowest metallicities).  

In this work, we investigated which avenue is more plausible between imposing the pair-instability at all metallicities and invoking gas accretion at the lowest metallicities to form massive BHs or relax the pair-instability at low metallicities and allow for the formation of massive BHs directly from the stellar collapse. 
Our key findings in this work can be summarized as follows:
\begin{itemize}
 \item The primary BH mass function can be explained by either invoking a high formation efficiency of Pop III black holes that have masses above the pair-instability limit at formation or by assuming about 10\% of BHs with masses below pair-instability limit at low metallicities to grow their mass through gas accretion from the ambient medium. 
 \item We find that in order to explain the primary BH mass function without invoking pair-instability physics at low metallicity would lead to a shallower relation between initial-final relation between the stellar mass and BH mass at the lowest metallicities. 
 \item While we find that a 10\% efficiency for BHs to grow their mass through gas accretion suffices to explain the observed primary BH mass function, higher values are plausible in principle while a lower efficiency for gas accretion mechanism would fail to match the data. 
\end{itemize}

To provide a global perspective on the derived merger rates as a function of redshift combined with formation efficiency of BHs at different metallicity bins, we compute the total energy density in GWs due to BH mergers of different metallicities. The left panel of Figure \ref{fig_discuss} shows he predicted redshift evolution of the BBH merger rate assuming 10\% of the BBHs at low metallicities double their mass before merging due to ambient gas accretion in dark matter halos at high redshifts. The large impact of the extremely high inferred formation efficiency of the BBHs at low metallicity is quite clear. Even though star formation at low metallicities is only relevant at the highest redshifts and for a rather brief cosmic period, the BBH merger rate in the local universe is comparable to those formed at higher metallicities in the later cosmic epochs. 
The right panel shows how $\Omega_{\rm GW}$ evolves with redshift for different metallicity bins. 
Here we assume that the secondary component in a BBH system follows a probability distribution of $p(m_2)=c$ bounded from below by 5 $\msun$. At each metallicity bin, we compute the radiated energy from the average BBH mass and the remnant BH mass. Our modeling shows that the overall $\Omega_{\rm GW}$ is about $1.5\times10^{-10}$ with most of the overall contribution being dominated by the BBHs formed at low metallicities. 
Our estimate is in line with the original estimate of \citet{Fukugita2004ApJ} of $\Omega_{\rm GW}\approx 10^{-9}$ and the current non-detection of the SGWB at LIGO frequencies.


\section{Acknowledgement}
MTS is supported by Dean’s Competitive Fund for Promising Scholarship at the Faculty of Arts \& Sciences of Harvard University.
E. R-R is supported by the  Heising-Simons Foundation, and NSF (AST-1911206 and AST-1852393).

\bibliographystyle{yahapj}
\bibliography{the_entire_lib}

\end{document}